\documentclass[10pt,conference,letterpaper]{IEEEtran}
\usepackage{times}        
\usepackage{helvet}       
\usepackage{courier}      

\usepackage[hyphens]{url} 
\usepackage{xspace}
\usepackage{balance}
\usepackage{cite}
\usepackage{textcomp}
\usepackage{epsfig} 
\usepackage{epstopdf} 
\usepackage{subfigure}
\usepackage{multirow}
\usepackage{xcolor} 
\usepackage{enumerate} 
\usepackage{moresize} 

\usepackage{tikz}
\usetikzlibrary{shapes}

\usepackage{gensymb}
\usepackage{amssymb}
\usepackage{pifont}

\def\pname{Gauth\xspace}

\def\ptitle{Hands-Free One-Time and Continuous Authentication Using Glass 
Wearable Devices}

\def\pkeywords{Authentication, HCI, wearable devices}

\newcommand*\circled[1]{\raisebox{.5pt}{\textcircled{\raisebox{-.2pt}{{\footnotesize#1}}}}}

\newcommand{\diamonded}[1]{\raisebox{-0.4em}{\tikz\node[diamond,draw,scale=0.3] {\Huge#1};}}

\newcommand{\squared}[1]{\raisebox{-0.2em}{\tikz\node[rectangle,draw,scale=0.65] {\Large#1};}}

\newcommand\crule[3][black]{\textcolor{#1}{\rule{#2}{#3}}}

\newcommand{\mycircle}[2]{\tikz\draw[#1,fill=#1] (0,0) circle (#2);}

\newcommand{\point}{\mycircle{black}{0.5ex}}

\definecolor{qr20}{HTML}{00477F}
\definecolor{qr50}{HTML}{4C88BE}
\definecolor{qr120}{HTML}{8DC3E9}

\def\ie{{i.e.,}\xspace}
\def\eg{{e.g.,}\xspace}
\def\etc{{etc.}\xspace}

\newif \ifcomments

\ifcomments
    \newcommand{\gp}[1]{{[[\textcolor{magenta}{GP: #1}]]}}
    \newcommand{\dd}[1]{{[[\textcolor{blue}{DD: #1}]]}}
\else
    \newcommand{\gp}[1]{}
    \newcommand{\dd}[1]{}
\fi

\newcommand{\comment}[1]{}

\newcommand{\surveyres}[1]{{\xspace\emph{\small #1}}}

\begin{document}

\author{
    \IEEEauthorblockN{Dimitrios Damopoulos \hspace{1em} Georgios Portokalidis}
    \IEEEauthorblockA{Stevens Institute of Technology\\ Hoboken, NJ, USA\\
    Email: \{ddamop,gportoka\}@stevens.edu}
}
\title{\ptitle}

\maketitle

\begin{abstract}

Users with limited use of their hands, such as people suffering from 
disabilities of the arm, shoulder, and hand (DASH), face challenges when 
authenticating with computer terminals, specially with publicly accessible 
terminals such as ATMs. When authentication through a password or PIN is 
possible, these users have an additional reason to choose a short ``easy'' 
combination due to the difficulties involved with entering lengthy convoluted 
passwords. Access tokens, like smartcards, can assist, however, they require 
that the user can physically handle such a token and custom reading sensors 
need to be installed in access terminals. Similar authentication challenges
are also present in environments where users need to frequently authenticate
and log out or require hands-free authentication, like in hospitals.

A new glass wearable device was recently introduced by Google and it was 
immediately welcomed by groups of users, such as the ones described above,  
as Google Glass allows them to perform actions, like taking a photo, using 
only verbal commands.
This paper investigates whether glass wearable devices can be used to 
authenticate users, both to grant access (one-time) and to maintain access 
(continuous), in similar hands-free fashion. We do so by designing and 
implementing \pname, a system that enables users to authenticate with a 
service simply by issuing a voice command, while facing the computer terminal 
they are going to use to access the service. To achieve this goal, we create 
a physical communication channel from the terminal to the device using 
machine readable visual codes, like QR codes, and utilize the device's 
network adapter to communicate directly with a service. More importantly, we 
continuously authenticate the user accessing the terminal, exploiting the 
fact that a user operating a terminal is most likely facing it most of the 
time. We periodically issue authentication challenges, which are displayed as 
a QR code on the terminal, that cause the glass device to re-authenticate the 
user with an appropriate response.
We evaluate our system to determine the technical limits of our approach. We 
show that even with the relatively low-resolution camera of the current 
Google Glass prototype, QR codes can be consistently processed correctly (with an average accuracy of 87.8\%), and continuous authentication, while strenuous to the 
battery, is feasible. Finally, we perform a small user study involving 
students to demonstrate the benefits our approach. We found that 
authenticating using \pname takes on average 1.63 seconds, while using 
username/password credentials takes 3.85 seconds and varies greatly depending 
on the computer-literacy level of the user. 

\end{abstract}

\section{Introduction}

Google recently introduced a glass wearable device, named Google Glass, 
that received warm welcome from certain groups of users for various 
reasons. People suffering from disabilities that diminish their ability to use
their hands welcomed it because it allowed them to perform tasks by issuing 
verbal commands alone~\cite{boon2disabled}. It also found use from medical 
doctors, who used it 
to access critical information during an operation without having to shift 
their attention away from the task at hand~\cite{glassOR}. This paper 
investigates whether glass wearable devices, like Google Glass, can be also 
used to improve the authentication process, both in 
terms of usability and security. We are focusing on two aspects: initially 
authenticating a user with a system (one-time authentication), and 
accurately determining when a legitimate user has stopped using a system by 
continuously authenticating him.

We focus on two groups of users to guide us in our research
(discussed in detail in Sec.~\ref{sec:examples}). Briefly, we first examine the 
challenges faced by people with disabilities in authenticating with 
existing systems, and in particular, with public terminals and 
devices shared by multiple users, where 
solutions like password managers and never logging out are \emph{impractical}.
As expected, these users suffer both from usability and security issues, which
is also confirmed by a recent study~\cite{disabilitiesauth:ares12} that showed
that they usually select 
simpler passwords and they are more prone to snooping~\cite{glass_snoopers}
when entering their passwords. Second, we consider the authentication 
challenges in environments like hospitals, where users need to frequently 
switch between terminals. Users try to avoid having to repeatedly enter their
passwords, or are absentminded, resulting in them leaving terminals without 
logging out with whatever that entails for security.

Some of the above issues can be overcome by using access tokens, 
like smartcards, instead of passwords. Tokens are more secure and potentially
more usable, than passwords, however, the user still needs to be able to 
physically handle one with his hands, and they do not provide a solution to
terminals being abandoned by users without logging out.
They also require that custom reading sensors are installed on terminals,
which can be costly and hard to maintain for services with large numbers of 
terminals.

The usual solution for preventing unauthorized access to abandoned 
terminals is to use timeouts, \ie logging out a user from a terminal or 
locking it after a period of inactivity. However, timeouts frequently lead to 
users being accidentally logged out. For example, consider a physician 
studying a patient's file without interacting with terminal, which cannot 
distinguish if the user is still there. Proximity sensors have been used to 
address this uncertainty, however, they also frequently make errors, like 
logging out active users and authenticating passing-by 
users~\cite{proximity_problems}.
 
We attempt to provide a solution to the above problems by proposing a new 
design that incorporates glass wearable devices to offer hands-free 
one-time and continuous authentication. Our approach, called \pname [goth], 
transforms glass devices to authentication tokens that can authenticate the 
user with supporting terminals just by facing them and issuing a voice 
command. We exploit the front-facing camera of glass devices to form an 
optical communication channel between a terminal and a device. 
Specifically, we use a visual code, such as a QR code, to transfer 
information regarding the terminal being viewed to the device, and use it 
to authenticate directly with the service behind the terminal. 

To provide continuous authentication, we exploit the observation that users 
operating a terminal will be most likely facing its screen. This allows us 
to use the same optical channel between device and terminal to continuously 
authenticate the user in front of it. The terminal periodically issues
re-authentication challenges, by displaying a QR code, which is captured by 
the glass, which, in turn, re-authenticates with the service. Our approach,
enables us 
to quickly identify when a user moves away from a terminal to lock it and 
eventually log him out. For example, during the evaluation of our prototype 
we used a period as short as five seconds. Intrinsically, \pname will
\emph{not} accidentally lock a terminal when a user is \emph{just studying} 
something on a terminal without interacting with it. Even if a user's 
terminal is erroneously locked, because he has to focus his attention on 
something or someone else, displaying a re-authentication QR code on the 
locked screen ensures that he will be transparently re-admitted when he 
returns to the terminal.

We implemented a prototype of \pname on Google Glass, which we evaluated by 
using it to authenticate with an e-mail service, which we also developed. 
We also performed a small user study, after obtaining IRB approval from our 
institution, to obtain real data from users operating our system. The study was
accompanied by a short exit survey to gauge user attitude toward wearable 
devices and \pname in particular.

Employing QR codes in authentication systems is not a new idea.
The last four years alone various works have proposed authentication 
systems combining QR codes and one-time passwords. They include systems for 
online banking~\cite{YoungSil2010, Young2010}, health 
systems~\cite{Mungyu2013}, access control~\cite{Kao2011}, 
web applications~\cite{Dodson2012}, 
personal smartphones~\cite{Starnberger2009, Kyeongwon2011, Roalter2013, 
Kwiecien2014}, and voting systems~\cite{Falkner2014}. 
None of these proposals offers hands-free or continuous authentication to 
users. We have also seen applications using Google Glass for two-factor 
authentication~\cite{glass2fa} (2FA), where one-time passwords (OTP) are 
displayed on its small screen for the user to type in a terminal after 
supplying his username/password. During our user study, we evaluated this 
scenario and our
results indicate that users find it harder to use, than using a smartphone 
for the same purpose. We also recently became aware of a company that is
advertising 2FA for desktops and laptops that utilizes Google Glass to 
authenticate a user with his PC~\cite{saaspass}. We do not have sufficient 
information to compare \pname with this approach, however, it seems to 
differ in many aspects from our work. The most obvious being no continuous 
authentication support and a lack of a general protocol. 

To summarize the contributions of this paper are:
\begin{itemize}
    \item We propose a new method that incorporates glass wearable devices to 
    provide hands-free authentication that:
    \begin{itemize}
        \item can authenticate users faster and more securely than passwords
        
        \item continuously authenticates users, promptly securing 
        unattended terminals from unauthorized access
    \end{itemize}

    \item We design a protocol for performing hands-free one-time and 
    continuous authentication between glass wearable devices and services

    \item We implement \pname, a prototype of our proposal

    \item We perform a thorough evaluation of our proposal through a set of 
    benchmarks and a small user study that demonstrates the feasibility of 
    the approach and its benefits. Our results show that \pname authentication
    is faster than passwords and operates consistently even when using small
    QR codes
    
    \item We evaluate the potential of using \pname for two-factor authentication\gp{and?}
\end{itemize}

The rest of this paper is organized as follows. Section~\ref{sec:examples} 
discusses in detail two examples that motivate and put our work in context. 
In Sec.~\ref{sec:model} we present the model of our system, including 
assumptions and our threat model. An overview of the system's operation is 
given in Sec.~\ref{sec:overview}, while the protocol used for 
authentication protocol is presented in Sec.~\ref{sec:protocol}. 
Implementation details are in Sec.~\ref{sec:implementation}. We evaluate 
\pname and present the results of our user stude in Sec.~\ref{sec:eval}. 
Related work is discussed in Sec.~\ref{sec:related}. We discuss potential 
issues, limitations, and future work in Sec.~\ref{sec:discussion}, and 
conclude in Sec.~\ref{sec:conclusions}.

\section{Motivating Examples}
\label{sec:examples}

While \pname can be beneficial for many types of users and organizations, we 
present two scenarios that we believe clearly illustrate the problems our 
solutions aims to address.

\gp{Mention that the device is a general purpose one and in both these
 scenarios users have also other uses of the device.}

\subsection{Authentication for Users with Disabilities}
\label{ssec:disabilities}

Accessing a computer terminal is a routine process for most computer-literate 
users today. Many of us operate a variety of different devices during a 
single day, ranging from personal smartphones, tablets, and laptops to 
specialized computer terminals used for banking, buying transportation 
tickets, and so on. Users with disabilities frequently face various 
challenges when operating such terminals. For instance, people with limited 
use of their hands, such as people suffering from DASH and Parkinson's, may 
require bigger buttons (on screen or physical), while visually-impaired users 
may require high-contrast displays. Fortunately, modern terminals are 
frequently equipped to assist such users. Concurrently, entities like the 
European Union and the United States government have introduced legislature 
pushing for further improvements in universal accessibility~\cite{section508, 
eu_disability_strategy}.

However, performing authentication remains challenging for many groups of 
users~\cite{disabilitiesauth:ares12}. Authentication is required to 
identify the user and to perform access control.
It is prominently performed using a password or PIN number that needs to be 
memorized by the user. Alternatively, machine-generated codes can be 
generated by specialized hardware (\eg one-time-password generators) or 
sent to the user from the authenticating service via a message to his 
personal mobile device.

A past study~\cite{disabilitiesauth:ares12} has shown that users with 
disabilities take longer to authenticate when using passwords. Besides 
degrading their experience, slower typing may also allow third parties to 
observe the password being typed, specially in public 
terminals~\cite{glass_snoopers}. At the same time, users tend to choose 
simpler, easier to guess passwords to facilitate their entry. For example,
they do not choose characters that require using multiple keys in a 
conventional keyboard, like symbols, or they simply choose shorter passwords.
Easy to guess passwords are vulnerable to offline attacks~\cite{cloudcracker,
gpu-cracking, ieee-leak, twitter-leak, sony-leak}, while not constraining 
failed authentication attempts may also enable online attacks~\cite{hydra}. 
The problem is only exacerbated when using space-restricted keyboards, such as
the ones on smartphones and tablets, as another recent study with typical 
users found~\cite{password_entry_method:arxiv14}. 

When using personal devices to access information, password managers can 
alleviate such issues, as the password needs to be only typed once. On 
tablets and smartphones, it is also common that a password is only entered 
the first time an application is started and a session cookie is 
established and used thereafter to authenticate the user. These solutions 
are much harder, or even impossible, to apply when considering public 
terminals or terminals and devices shared between users (\eg in a workplace).

Another approach is to distribute authentication tokens to users, such as 
smartcards and USB keys. These tokens can be presented to terminals for 
authentication. While using tokens can be safer and faster than passwords, 
it has the following disadvantages. First, custom readers may need to be 
installed on many terminals. For example, RFID readers for contactless 
smartcards. Second, the user still needs to be able to handle the 
particular token with their hands, which could be problematic for some 
groups of users.

Our approach aims to allow users with disabilities to authenticate without 
the need to use their hands. We are targeting users that can still operate 
a terminal, so they do have some use of their upper extremities, and their 
vision is sufficient. Our goal is to transform glass wearable devices to a 
virtual key chain that can hold keys for different services, can be 
easily utilized by users with disabilities, and has a low adoption threshold 
by services.


\subsection{Continuous Authentication for Physicians and Nurses in Hospitals}

Physicians and nurses use both portable devices and fixed terminals located 
within hospitals. The fixed terminals can be in public spaces (\eg spaces 
accessible to patients and visitors) and are usually shared by many 
different users (\ie the hospital staff). Consequently, authentication is 
required for two reasons: first, to prevent unauthorized use of the 
terminals and, second, to present each user with their own personal 
environment and data that allows them to be more efficient when moving 
between terminals.

When users leave their terminal, it is important that they 
log out. However, they frequently neglect to do so, because they are forgetful
or in a hurry. Some times, users request to log out and immediately move away 
from the terminal, but their request requires confirmation to complete because
they have left unsaved work, which would be otherwise lost. In other cases,
the decision to leave a terminal without logging out is intentional and aims
to save time, as physicians frequently perceive authentication as being 
slow~\cite{healthcareIT_problems}, and a physician returning to a logged out 
terminal would have to authenticate again. It is obvious that this behavior 
can leave terminals exposed to unauthorized users, but this is not the only 
problem. Past studies reported that physician accessing terminals with 
another user already logged in has resulted in medication 
errors~\cite{medication_errors}, because physicians entered data in the wrong 
patient's sheet.

The most common solution to the issue is to use timeouts, \ie logging out a 
user from a terminal or locking it after a period of inactivity. Picking 
the right value is not an easy task though. Using a short timeout may lead 
to erroneously logging out users that are still using a terminal but do not 
interact with it. For instance, a physician studying a patient's file or 
momentarily discussing with a patient cannot be distinguished from him 
walking away from the terminal entirely~\cite{healthcareIT_problems}. On the 
other hand, picking a long timeout means that the same problems discussed 
above are still possible. An alternative to timeouts is using proximity 
sensors. However, they frequently erroneously log out active users or 
authenticate passing-by users, leading to frustration and hacks like 
disabling them by putting a Styrofoam cup over the 
sensor~\cite{proximity_problems}.

Our approach aims to enable systems to continuously authenticate the user 
operating a terminal and avoiding erroneously logging him out or locking 
the terminal. Our system allows for checks that determine whether a user is 
still looking at the terminal before taking action, while even in the case 
where a terminal is locked, because the user has briefly moved away, it will 
transparently unlock it when he returns and faces the terminal to resume 
his task.

\section{System Model}
\label{sec:model}

\subsection{Assumptions and Requirements}

\subsubsection{Device Requirements}
Our approach requires a head-mounted device featuring a front facing camera
that can take static pictures from the point-of-view of the user. While a 
high-definition camera is not necessary higher resolution cameras will allow 
us to use smaller visual codes on displays. A screen is not required, however,
if present, it can be used to provide feedback to the user (\eg the name 
of the service he is authenticating with) and for locking the device (see 
Sec.~\ref{ssec:security}). Sensors such as a microphone and a gyroscope are 
also necessary for receiving voice commands and turning on the device, 
respectively. The device should have a CPU capable of decoding QR codes, 
performing simple cryptographic operations, and running simple algorithms 
for identifying simple voice commands. Finally, it should also include a 
network adapter (\eg WiFi) that would allow it to connect to the 
authentication service being used and sufficient storage to store a user's 
credentials for all his \pname-enabled services.

\subsubsection{Device-User Association}
A device can be associated with a single or multiple users. In the first 
case, the device is considered personal and identifies the user itself, 
while in the latter the device is associated with a particular role or set 
of permissions. For example, this can be the case in hospitals or military 
compounds where the device essentially authorizes the user to access a set 
of terminals and services. Alternatively, shared devices may be assigned to 
particular users when checked out for use (\eg when a nurse's shift starts).
Users may also be required to supply a PIN before using a device, while 
biometrics can be used for the same purpose~\cite{whowearsit:mobisys14}. 

\subsubsection{Device-Service Association}\label{ssec:ds_assoc}
A \pname device can authenticate a user with multiple services. We assume 
that before it is used, it has been properly set up to identify him with a 
service, after authenticating using alternate means, like with a username 
and password. Necessary information to complete the association can be 
shared by scanning a QR code with the glass device, similarly to how 
Google's one-time password generator (OTP) is 
initialized~\cite{google-authenticator}, or by using another available pairing
method~\cite{pairing:jpmc09}. A user could also physically present the device 
along with identification to a service (\eg the branch of a bank) to perform 
the association. 


\subsubsection{Deployment Requirements for Services and Terminals}
Supporting \pname-based authentication does not require any new sensors or
hardware to be installed at terminals, assuming that they include a screen 
that can display graphics, like a QR code. If the user interface (UI) presented
on a terminal is \emph{service driven}, then no software changes are 
required on terminals either. For example, in the case of terminals receiving 
HTML content from a service and thin clients~\cite{hospital_thin_clients}. In
the opposite case, if the UI is \emph{terminal driven}, software modifications
on the terminal are required to generate and display visual codes.


\subsection{Threat Model}

\subsubsection{Adversaries}
The most significant threat faced by \pname users is theft of their 
wearable device. An attacker obtaining the device and knowing the services and
terminals it has been associated with, can use it to gain access to services,
obtain personal information, perform financial transactions, \etc This 
is akin to a thief stealing someone's home key. The device can be 
protected using a PIN or biometrics and the information on the device can 
be encrypted. In this case, the attacker would have to brute force the PIN 
or mimic the biometric used to unlock the device, before being able to use 
it. In the case of continuous authentication, an adversary could attempt to
use a terminal, left by a user without logging out, before \pname detects 
that the correct user is no longer behind the terminal. Increasing the 
frequency that a service requests re-authentication can reduce this window 
of vulnerability. In Sec.~\ref{sec:eval}, we evaluate the effects of using 
different re-authentication frequencies and how it affects the window of 
opportunity for an attacker.

Finally, an adversary could launch an online brute-force attack against a 
\pname service, because the authentication service needs to be accessible 
to glass devices, which in certain contexts implies that it needs to be 
accessible over the Internet. However, the probability of such an attack 
succeeding is extremely low, as the attacker cannot observe the user's ID or 
his authentication credentials, and such an attack would be easily detected by
the service. An attacker can, though, obtain the information contained in the 
visual codes displayed on the terminal, which are considered public, and 
identify the terminal and service to the device.

\subsubsection{Device Integrity}
Our approach has similar security characteristics to software security tokens
on smartphones (\eg Google authenticator~\cite{google-authenticator}).
As such, we assume that \pname software running on the wearable device, the 
operating system kernel, and hardware, have not been compromised. Since 
devices like Google Glass permit users to download apps from stores, we 
assume that the user may have installed one or more malicious apps, 
however, these apps cannot compromise \pname by reading or altering its 
functionality and data.

\subsubsection{Users}
We consider the users of the system benevolent, however, they may expend 
small effort to bypass the system, if it interferes with their work. For 
instance, they may attempt to trick the glass device, so that they remain 
continuously logged in, and a terminal never locks or logs them out.

\subsubsection{Man-in-The-Middle Attacks (MitM)}
MitM attacks between the terminal and the service cannot be prevented by 
\pname. Thus, we assume that terminals and services can mutually 
authenticate (\eg using TLS, certificates, and certificate authorities). A 
MitM can relay information between the terminal and service to enable the 
user to authenticate using \pname. However, our approach submit the user's 
credentials directly to the service, so they are not exposed to the attacker.
We assume that MitM attacks between the device and the service are 
not possible, because the device has been properly set up with the service's 
certificate, certificate authorities are used properly, \etc

\section{System Overview}
\label{sec:overview}

\pname transforms glass devices to authentication tokens that
can be used to authenticate the user with different services. This section 
describes our approach for performing hands-free \emph{one-time} and 
\emph{continuous} authentication.

\subsection{One-time Authentication}

\begin{figure}
    \centering
    \includegraphics[width=\linewidth]{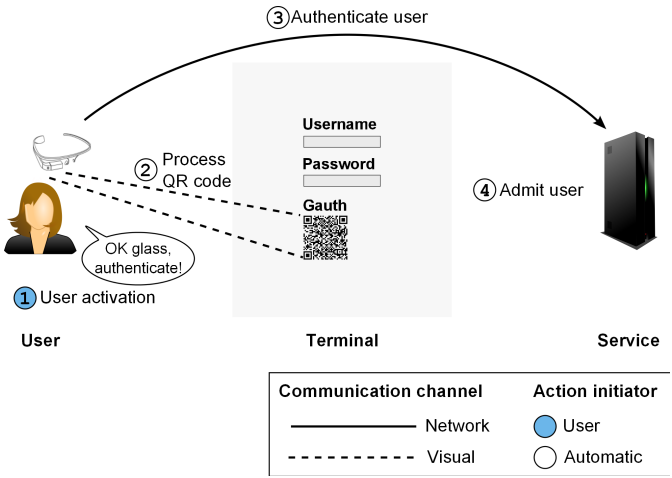}
    \caption{One-time, hands-free authentication with \textbf{\pname}.}
    \label{fig:one-time_overview}
\end{figure}

Initially authenticating with a system using \pname comprises of the steps 
depicted in Fig.~\ref{fig:one-time_overview}. \pname-enabled services display
a visual code, such as a QR code, on unused terminals, which can either 
replace the previously used authentication interface or coexist with
it. This visual code encodes identifiers for the service and terminal, but 
it can also include a random number (nonce), which is refreshed every time 
a new log-in dialog is presented, to produce QR codes that are different 
every time. Introducing randomness is necessary to ensure that a stored 
picture of an old QR code cannot be used to authenticate with a terminal. 

Authentication is explicitly triggered by the user using a voice command, 
such as ``OK glass, authenticate'' (step \circled{1}). Google Glass is 
equipped with a voice 
command system allowing developers to enable any application through voice. 
It can be programmed to accept new commands, and it does not need specific 
training to ``understand'' new voices. Its accuracy can drop, though, when 
the Glass is not connected to the Internet, as cloud resources are used to 
enhance the speech-to-text analysis. Hence, we can program specific 
keywords that will trigger authentication. As the device enters power-save 
mode when not used, the user may need to also perform a small 10\textdegree{} 
head tilt to wake up. This is enabled by the gyroscope sensor on the device.

The device, which has been previously set up to authenticate the user with the
service, takes a picture of what the user is looking at, scanning for a QR 
code (step~\circled{2}). If the device fails for some reason to find a 
code, it will retry, until a configurable threshold is reached, in which 
case it can provide visual or auditory feedback to the user. The 
information extracted from the QR code is used to identify whether the device
has been associated with the corresponding service. Once 
such a service is found, the device communicates with it (\eg over WiFi,
3G, \etc), presenting the user's credentials and the corresponding terminal 
identifier to authenticate the user (step~\circled{3}). Finally, the service 
verifies the credentials received and admits the user in the requested 
terminal (step~\circled{4}).

We see that while the user needs to initiate the process in step~\circled{1}, 
the rest of the process is automatic and transparent to the user.
Also, the user does not need to use his hands to type a password or PIN,
search for his smartcard, \etc

\subsection{Continuous Authentication}

\begin{figure}
    \centering
    \includegraphics[width=\linewidth]{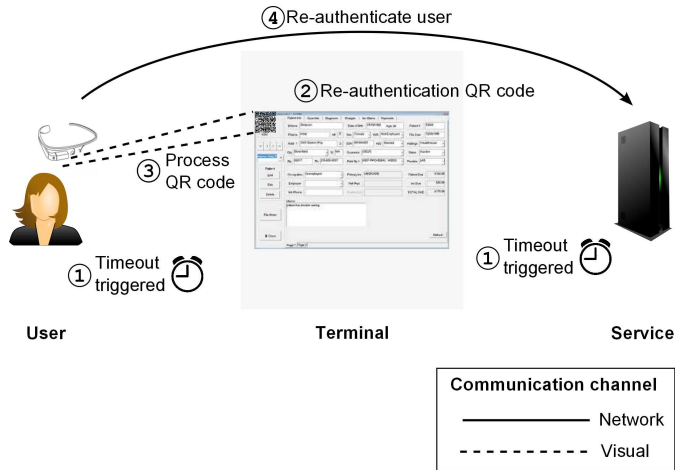}
    \caption{Continuous authentication with \textbf{\pname}.}
    \label{fig:continuous_overview}
\end{figure}

Devices enter continuous-authentication mode automatically after being 
signaled, as part of the acknowledgment returned to the device, when a user 
successfully authenticates initially. The process is based on the same 
principle as with one-time authentication, but uses timeouts to trigger, in 
this case, re-authentication. It begins with the device receiving 
acknowledgment that an authentication request has succeeded. At that point,
the service signals that it requires continuous authentication with a minimum
period of $T_{reauth}$ and synchronizes clocks with the device. This means 
that while the terminal is being used, the service expects a re-authentication
request from the device every $T_{reauth}$ seconds. 

The body of the continuous-authentication loop that follows is shown in 
Fig.~\ref{fig:continuous_overview}. When $T_{reauth}$ seconds elapse 
(step~\circled{1}), a new re-authentication challenge, encoded as a QR code,
is displayed on the terminal (step~\circled{2}). The QR-code encodes similar 
information as before and can be superimposed on the content in a manner 
that is non-disruptive. For instance, a QR code can be placed near the edge 
of the terminal. Obviously, a user interface that takes into account \pname 
could reserve some space for this purpose. As a rule of thumb, we try to 
use the smallest QR code possible to reduce interference with the user (in 
Sec.~\ref{ssec:qrcodes} we evaluate different QR-code sizes).

The device periodically takes snapshots scanning for re-authentication codes.
Since, essentially, the device has synced with the service, it will 
also automatically activate after $T_{reauth}$ seconds and start scanning for
a QR code (step~\circled{1}). As long as the user is operating the terminal 
and utilizing the 
screen, the device will eventually identify such a code. It will process it 
and re-authenticate the user by re-supplying his credentials, as well as 
the nonce extracted from the code, to the service (step~\circled{3}). 
Finally, the service verifies the user's credentials and nonce, and the code
is removed from the display. 

If an authentication token is not sent in a certain amount of time, after 
the challenge was initiated by the service, it can choose to lock the 
terminal or completely log out the user. Note that if the user has recently 
interacted with the terminal, the service could decide to be more 
permissive and offer more time for re-authentication. If the terminal is 
locked, the lock screen should display the last QR-code challenge made by 
the service to ensure that the user can quickly resume. This way, when the 
user returns his gaze to the terminal, he will be transparently
re-authenticated and the terminal unlocked. \emph{Note that while the device 
is in continuous-authentication mode, it will only issue authentication 
requests for the terminal it originally authenticated the user with.} So a 
user will \emph{not} accidentally authenticate with other locked terminals.

It may be desirable that a terminal remains unlocked without the user 
actively interacting with it. For example, a physician may be using a 
terminal to show information to a patient without at the same time looking 
at the terminal. In this case, the user would face a similar problem as to 
when only employing a timeout for determining inactivity, however, the QR 
code challenge could serve as queue for him to face the screen or interact 
with the terminal to prevent locking.

Finally, while a device is in continuous-authentication mode a visual 
marker can be shown on its display to provide feedback. The mode can be 
exited directly through a user request, which will also terminate the 
user's session with the associated terminal, or indirectly by the service 
signaling the device when the user logs out using the terminal's user 
interface.

\section{Authentication Protocol}
\label{sec:protocol}



\begin{table}
    \centering
    \caption{Keys and identifiers shared between protocol parties.}
    \label{tab:shared}

    \begin{tabular}{lccc}
                        & {\bf Device} & {\bf Terminal} & {\bf Service} \\
    \hline
    Used ID ($UID$)                 & \point    &           & \point\\
    User OTP Key ($K_{U}$)          & \point    &           & \point\\
    Service authentication URI      & \point    &           & \point\\
    Service ID ($SID$)              & \point    & \point    & \point\\
    Service certificate ($CERT_{S}$)& \point    & \point    & \point\\
    Nonce OTP Key ($K_{N}$)         &           & \point    & \point\\
    Terminal ID ($TID$)             &           & \point    & \point\\
    \hline
    \end{tabular}
\end{table}

\subsection{Preliminaries}
\label{ssec:proto_prelim}

The protocol used by \pname relies on the transport layer security (TLS)
protocol~\cite{tls} and one-time passwords (OTP). TLS is used to ensure the 
confidentiality of communications and enables the user to authenticate the 
service. To securely employ TLS between the user and the service, the user 
has already obtained the certificate of the service $CERT_{S}$ during 
device-service association (Sec.~\ref{ssec:ds_assoc}), which also includes
the service identifier ($SID$) the Internet location of the service in the 
form of a uniform resource identifier (URI). 

To authenticate the user 
to the service, we utilize time-based OTPs~\cite{totp} (TOTP) because they 
allow us to authenticate by sending a single message to the service. To 
utilize them, a shared key $K_{U}$ is initially exchanged between the 
service and the user. Using $K_{U}$ and the current time, the device can 
generate an $OTP$ that can be verified by the service. This is similar to 
OTP passwords generated by google-authenticator~\cite{google-authenticator}. 
Using the $CERT_{S}$ and an $OTP$, we achieve mutual authentication of 
user and service. 

The protocol also utilizes a $Nonce$, which is used to ensure that older, 
stale QR codes cannot be used to abuse the system. We generate it in two 
ways, depending on the party that drives the UI. If the terminal UI is 
service driven, then a random number is generated by the service. On the 
other hand, if the UI is terminal driven, then we use a \emph{HMAC-based 
OTP}~\cite{hotp} (HOTP) generator. This is done to enable the terminal to 
generate a nonce that can be checked by the authenticating service without 
the need to interact with it. This requires a shared key $K_{N}$ to also be 
shared between the terminal and the service.

The service certificate (including service ID and URI), along with the user 
ID and the shared key for generating user OTPs compose a 3-tuple $(CERT_{S}, 
UID, K_{U})$ that uniquely identifies an account the user possesses with 
a service. A summary of the keys and other information share between the 
three parties of this protocol is shown in Tab.~\ref{tab:shared}.

While the communication channel between the terminal and the service is not 
the focus of this work, we assume that it is also secure and, for 
simplicity, that it is over TLS.

\subsection{Protocol Steps}

\begin{figure}
    \centering
    \includegraphics[width=\linewidth]{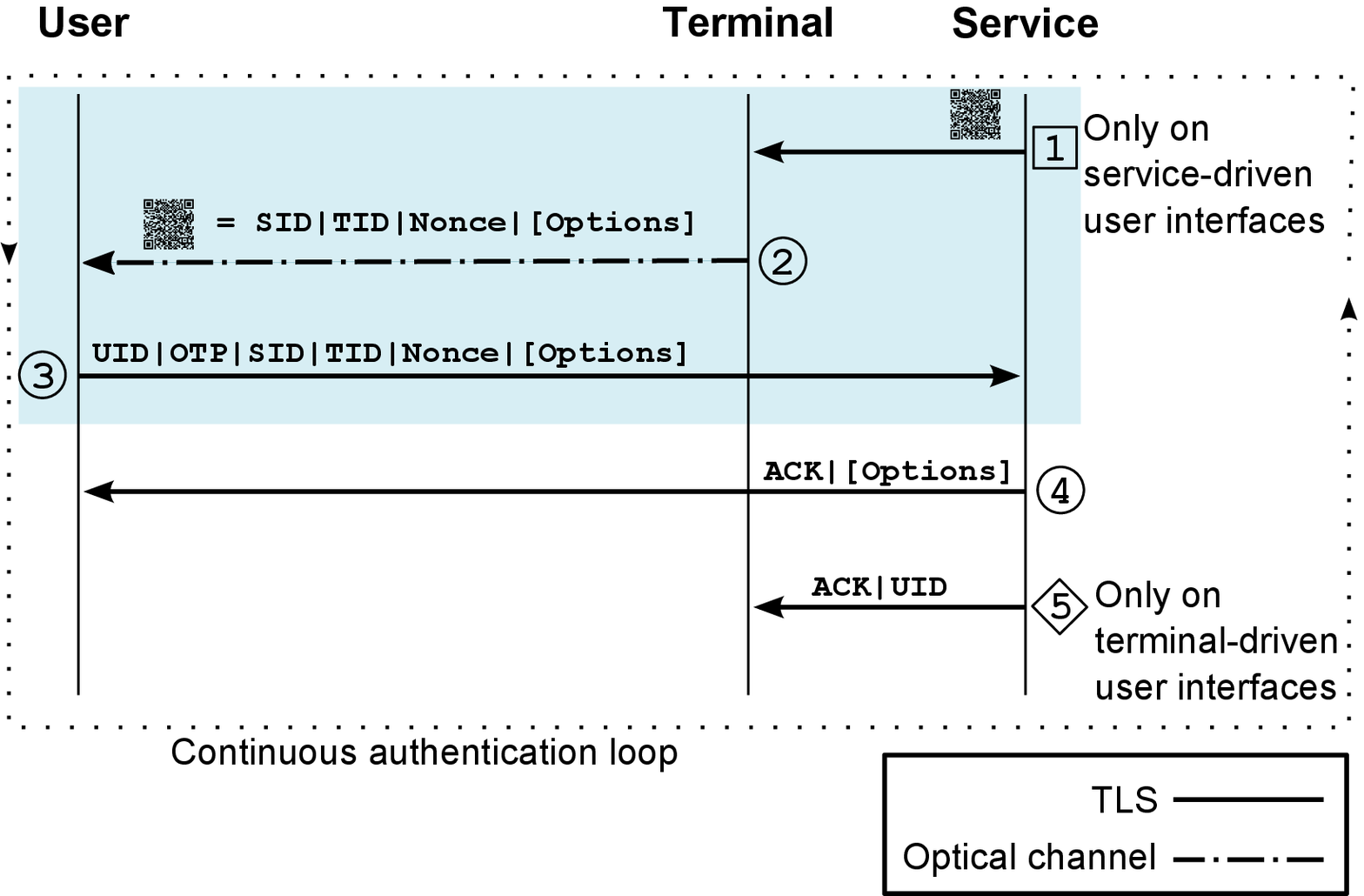}
    \caption{\pname authentication protocol.}
    \label{fig:auth_protocol}
\end{figure}

The protocol consists of five different steps illustrated in 
Fig.~\ref{fig:auth_protocol}. However, only \emph{four} steps are actually 
utilized during authentication. This is because step~\squared{1} is only 
performed when the UI is service driven and step~\diamonded{5} when it is 
terminal driven. In the first case, the service can manipulate the UI shown 
on the terminal to display a QR code (\eg by adding an image in an HTML page),
so the first step involves encoding the information into a QR code and 
displaying it on the terminal. In the second case, the software running on 
the terminal is the one generating the QR code.

Step~\circled{2} is performed over an optical channel formed between a 
terminal's screen displaying the visual code and the camera on the wearable 
device. The information sent to the device through the QR code includes the 
$SID$, $TID$, $Nonce$, and a configurable number of options.
The wearable device reads the QR code and decodes it to retrieve the 
information encoded. We rely on the intrinsic error correction in QR codes 
to ensure that the information is not corrupted.

Once the device decodes the QR code, it creates a secure TLS connection 
with the remote service using the stored URI (step~\circled{3}). Since the 
device holds the service's certificate, it can verify that it is 
communicating with the correct party. To authenticate the user, the device 
generates a new OTP using $K_{U}$ and the current time, and sends it to the 
service along with the information shown in Fig.~\ref{fig:auth_protocol}.

The remote service can use the information to verify the identify of the 
user in step~\circled{4}. In particular, it uses $UID$ to locate the user's 
$K_{U}$ and verify the received OTP. The service also verifies that the 
received $Nonce$ matches what is expected for $TID$. As discussed in
Sec.~\ref{ssec:proto_prelim}, the nonce can be a randomly generated number or
an HOTP. After, the service then responds to the device with a 
positive or negative acknowledgment (step~\circled{4}), which may also 
include additional options. For example, it can flag the glass that the 
terminal requires continuous authentication over \pname and send the
re-authentication period $T_{reauth}$.

If the UI is terminal driven, the service notifies the terminal of the 
outcome of the authentication process in step~\circled{5}. If it was 
successful, the $UID$ of the user is also sent to the terminal.

During continuous authentication the protocol is repeated, as shown in
Fig.~\ref{fig:auth_protocol}, until the user logs out or his session is 
terminated by the service or terminal. Additionally, the options in 
steps~\circled{1}~and~\circled{3}, which are highlighted in 
Fig.~\ref{fig:auth_protocol}, are used synchronize the clocks of the 
device and the service (or terminal) to ensure that re-authentication can proceed 
successfully. Our approach is based on Lamport's timestamps
algorithm~\cite{Lamport:1978}, which is used to determine the order of events
in a distributed computer system.

\section{Implementation}
\label{sec:implementation}

\pname consists of primarily three components: an app residing on the glass 
device, an authentication layer running at the service back end, and 
potentially software running at the terminal. This section provides 
technical details about the implementation of these components and their 
interactions. For our prototype, we used a Google Glass device and built a 
\pname-enabled e-mail server and client in Java, representing the scenario 
that the UI is driven by the terminal, and the service cannot directly 
manipulate it. To implement the protocol we used a four-digit number for 
service IDs, and a six-digit nonce, which is generated by the terminal as an 
HOTP and also serves as its ID. In the options of the protocol, we add an 
11-digit timestamp to execute Lamport's algorithm~\cite{Lamport:1978}.

\subsection{\pname Google Glass App}
\label{ssec:gauth_app}

%

The \pname app is Glassware software that runs on a second generation 
Google Glass XE (Explorer Edition) device, equipped with the Android v4.4.2 
kernel and the XE 21.0 update. \pname was developed with the Glass 
Development Kit (GDK), an add-on to the Android SDK. It consists of four 
modules, implementing the following functionality.

\subsubsection{Voice Activation}
\label{sec:voice_commands}

The GDK's \textit{VoiceTrigger} framework allows the creation of a 
voice-driven interface via three types of speech recognition: application 
shortcut-type voice commands that enable the user to start an application, 
contextual voice commands that enable users to perform particular actions 
within the context of a running app, and free-form speech-to-text 
transformation for receiving textual input from users (\eg reciting an 
e-mail). To enable starting our app using a voice command, we register an 
intent filter using the 
\textit{com.google.android.glass.action.VOICE\_TRIGGER} action in the app's 
Android manifest. In our prototype we use the 
phrase ``OK Glass, Authenticate''. Additionally, we need to declare what 
resources need to be used by our app in file 
\textit{res/xml/Gauth\_voice\_trigger.xml}. In our case, we require 
to use the camera, network, and microphone.

\subsubsection{QR Code Scanning} 
 
To detect and decode QR codes we use the \emph{ZXing engine}. ZXing is an 
open source library able to scan and decode one and two-dimension barcodes 
using a smart device's camera. Unfortunately, the library available for 
Android depends on a remote service to decode QR codes, so we modified it 
to create a library that also includes ZXing's core image decoding 
library in order to quickly decode QR codes on Google Glass. This way we 
increase authentication speed and we avoid sharing QR codes with a 
third-party.

\subsubsection{One-time Password Generation} 
 
\pname implements time-based OTPs by integrating 
google-authenticator~\cite{google-authenticator}. After initially setting up 
the key shared with the server ($K_{U}$), requesting a new OTP is 
straightforward.

\subsubsection{TLS Support}
\label{sec:public_key}

\pname creates, store, and manages digital keys and certificates using the 
Java \textit{KeyStore} framework and the \textit{OpenSSL} library. By 
default, it stores and uses a list of trusted CA certificates, similarly to 
a browser or smartphone. In our case, we also include a base-64 encoded 
X.509 certificate for the e-mail service, which is part of our prototype. 
The communication with the remote service is implemented via the Java 
\textit{HttpsConnection} framework, which connects the device to the 
service back end over TLS.

\subsection{\pname Support for Service Back Ends}
\label{sec:remote_service}

\gp{Is the implementation small? aprox 600 lines}

Services need to support \pname either by design or through a middleware 
that enables third-party services to interface with \pname devices. In our 
prototype, we built such a middleware to add \pname support for an SMTP/POP3
e-mail service. We implemented it in Java and support two different 
network interfaces; the first one is capable of handling multiple \pname 
client connections from users that authenticate using the protocol 
described in Sec.~\ref{sec:protocol}, while, the second one, allows 
terminals, in this case e-mail clients, to be updated in real time with a 
user's authentication status. Note that this interface is required for UIs 
that are terminal driven, while UIs that can be directly modified by the 
service may not need it at all. 
In our prototype, communications are secured using HTTPS and implemented 
using the Java \textit{HttpsConnection} framework, however, other designs 
that rely on IPSec, or controlled environments like an Intranet or cloud 
infrastructure, can also be supported in a similar way with relatively 
little effort.



\subsection{Terminal}
\label{sec:terminal}

When a terminal UI cannot be directly modified by the service, but requires 
support from the software running on the terminal, some modifications are 
also required on its software. To test this scenario, we built a Java-based 
e-mail client to act as a thick terminal, which will generate the QR codes 
itself instead of the service. 
Encoding the protocol information in a QR code and and displaying it on the 
terminal is done using the \textit{ZXing} library. For creating an e-mail 
interface, we used the \textit{JavaMail} API~\cite{Javamail} which provides 
support for SMTP/POP3 server. Communication with the service is performed 
over TLS both for communicating with the \pname middleware, as well as for 
delivering the content/service users, in this case providing access to 
their e-mail accounts. To generate a nonce for the protocol, we use HOTPs 
created (again) by the google-authenticator 
library~\cite{google-authenticator}.

\section{Evaluation}
\label{sec:eval}

\begin{figure}
    \centering
    \includegraphics[width=\linewidth]{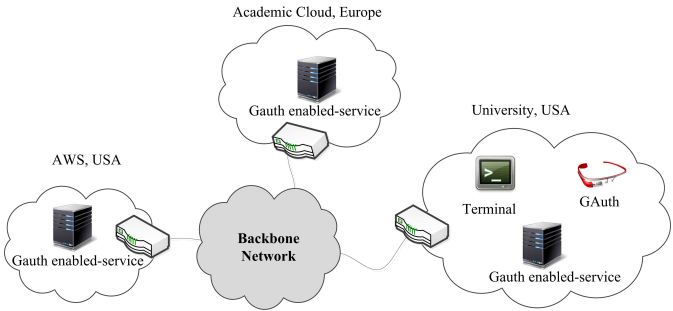}
    \caption{Testbed used during the evaluation.}
    \label{fig:testbed}
\end{figure}

​This section presents the results derived from the evaluation of our \pname 
prototype implementation. During our evaluation, we used the testbed 
topology drawn in Fig.~\ref{fig:testbed}. First, we deployed the terminal, 
service, and client in our own institutional network in the US. Then, we 
experimented by deploying the service to two cloud infrastructures. An 
academic cloud located in Europe and Amazon's AWS cloud in the US. 
Specifically, the testbed was comprised by the following elements:

\begin{itemize}
\item A second generation Google Glass, equipped with an OMAP 4430 SoC 
dual-core CPU, 2GB RAM, 5 MP Camera and a Wi--Fi 802.11b/g as the network 
interface, which runs \pname

\item One high--end laptop machine with Intel Core 2 Duo P8700 at 2.53 GHz, 
8 GB of 1,066 MHz DDR3 RAM and a 6 Mbps AirPort wireless network card, serving
as the terminal
 
\item One server with an Intel Xeon CPU at 3.2 GHz and 4 GB of RAM located in
the cloud service in Europe, acting as a remote service

\item One server with two Intel Xeon Processors operating at 2.5GHz with Turbo
up to 3.3GHz and 2 GB located in Amazon's EC2 cloud infrastructure, acting as
a remote service
\end{itemize}

We conduct a thorough evaluation of \pname in terms of performance, 
effectiveness, and efficiency using human subjects. Our aim is to 
demonstrate that \pname is faster than password-based authentication 
without compromising security. More precisely, we evaluated \pname's 
performance, QR-code readability, which affects the usability of our 
scheme, and battery consumption when doing continuous authentication. 
Finally, we performed a user study involving 20 students using \pname in 
different scenarios.

\subsection{Authentication Time}

\begin{table}
    \centering
    \caption{Authentication time in milliseconds. The results are from 
    performing 20 authentication runs for each service location.}
    \label{table:network_measurements}

    \begin{tabular}{l*{5}{c}r}
        \hline
        Remote service location & Mean & Min & Max & Standard deviation \\
        \hline
        Local  &  329 & 265 & 421 & 70  \\
        AWS (Oregon) & 523 & 451 & 609 & 134  \\
        Cloud (Europe) & 703 & 584 & 2115 & 476  \\
    \end{tabular}
\end{table}

Authentication time refers to the overall time required for a \pname user 
to be authenticated by the remote server. We calculate the mean time from 
the moment the user activates \pname via voice, until a complete 
authentication is performed. Authentication time includes all the 
necessary actions \pname requires for authentication as already described 
in Fig.~\ref{fig:auth_protocol}. Furthermore, to correctly evaluate
\pname's authentication time, we collect measurements from 20 authentication
runs performed for the three locations of our testbed, and we calculate the 
average.

Overall, the mean authentication time with \pname is 3.8~sec. This number 
includes 1.8~sec required to capture a picture using the auto-focus feature 
of the camera, followed by the QR-code decoding process 0.2~sec. Over and 
above that, it requires 0.4~sec to generate a new OTP, while the network 
communication part over HTTPS takes an average of 0.6~sec. To activate 
\pname via voice command requires an average of 1.7~sec.

This means that when doing continuous authentication with a period of 
5~sec, \pname is able to perform almost 12 incessant authentications within a 
minute. Of course, authentication time is affected by the network's quality 
and potentially geographical distance between the user and the remote 
service. Table~\ref{table:network_measurements} summarizes our results, 
including the mean, minimum, maximum, and standard deviation of our 
measurements. Note that our measurements incorporate the time required to 
establish an HTTPS connection with the remote service and end when \pname 
receives the authentication ACK, as described in 
Fig.~\ref{fig:auth_protocol} (\ie messages 3 to 5).


\subsection{Overhead}

\begin{figure}
    \centering
    \includegraphics[width=\linewidth]{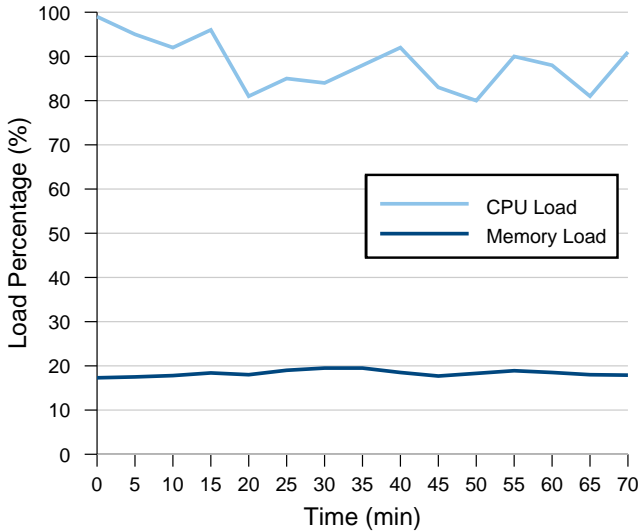}
    \caption{CPU and memory load while performing continuous authentication 
    with a period of $T = 5$ seconds. CPU load is higher than expected due to 
    an open issue with the Glass' camera framework~\cite{CPUload}.}
    \label{fig:cpu_mem}
\end{figure}

By overhead, we refer to the computational and memory resources \pname 
consumes. Wearable devices do not afford considerable CPU and memory 
resources, at least when compared to smartphone and tablets, so overhead 
can be decisive. Some pilot results on CPU and memory utilization for \pname
 are presented in Fig.\ref{fig:cpu_mem}. These results correspond to \pname 
performing continuous authentication with a period of $T = 5$ seconds. As 
observed from the figure, we monitored the application for about 70 minutes 
before the Glass' battery discharged completely. The maximum and minimum 
CPU consumption during this time period is 98\% and 80\% respectively, 
while the memory consumption was about 18\% -- 20\%. Overall, what we 
observe is that there is a significant increase in CPU usage during the 
authentication process. However, this happens due to an open issue with the 
Glass' camera framework~\cite{CPUload}. Additionally, we need to mention 
that the temperature of the device was notably higher during this 
experiment, and we even encountered overheating issues, receiving the error 
message \textit{``Glass must cool down to run smoothly''}, when we tested 
\pname with a smaller authentication period (T\textless 5). ​

\subsection{Battery consumption}

By default continuous authentication can be a power hungry operation, which 
can greatly affect battery-powered devices. In contrast to the official 
estimation from Google of ``one day battery life'', our experiments 
indicate that the Glass' battery usually drains faster. 

\begin{figure}
    \centering
    \includegraphics[width=\linewidth]{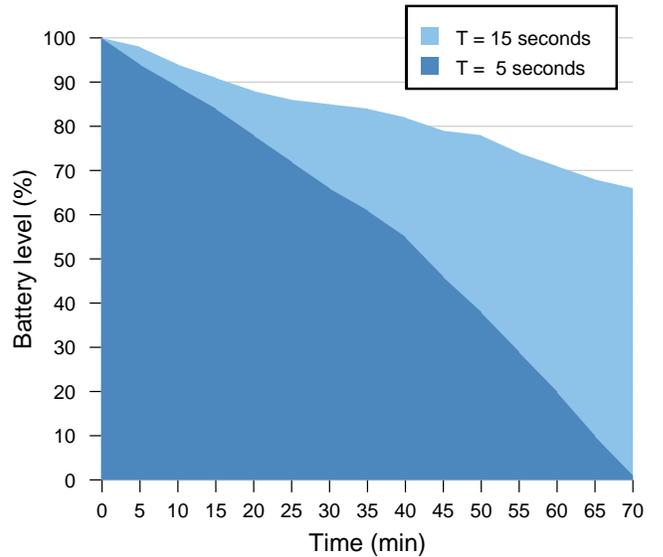}
    \caption{Battery consumption during continuous authentication. \pname is 
    configured to wake and take a picture every $T$ seconds.}
    \label{fig:battery}
\end{figure}

In an effort to balance the continuous authentication period against 
battery consumption, we measure the battery level every minute, while \pname
performs continuous authentication. Figure~\ref{fig:battery} illustrates 
the battery consumption during this experiment, when using a period of 5 
and 15 seconds.

With an authentication period of T=5 (sec), \pname is taking 2\% off the 
battery every minute, while with period of T=15 (sec) we gained 2/3 of more
energy. Practically, this means that within an hour, \pname is able to perform
720 authentications with the remote service for T=5 (sec).

In terms of completeness, we also measure 4 fundamental Glass operations for a 10 minute period. Basic operations, like stand-by, playing video or receiving and displaying GPS directions, can drain 2.5\%, and 6.2\% of the battery respectively, while video recording is in position to consume 31\% of the battery within just 10 minutes. ​

\subsection{QR-code Readability}
\label{ssec:qrcodes}

The key element of our scheme depends on the high readability level of the QR-code itself. A QR-code consists of black square dots --modules-- placed in a square grid on a white background, which can be read by a camera and processed using Reed--Solomon error correction until the image can be appropriately interpreted~\cite{QRCode_standart}.
Failing to correctly identify and decode a QR code means that the user will not be able to authenticate and would have retry. The code's physical size, the amount of information stored in it, the distance between the code and the scanning device, the position or the viewport of the camera, lightning, and, last but not least, the camera lens are some of the reasons that affect the readability of a QR-code. Thus, stressing both \pname and the QR-code to identify their limits was a primary task. However, in this evaluation, we do not evaluate factors like the lighting or the quality of the camera lens.

The QR-code evaluation was performed using the native 5 MP Google Glass camera using the auto--focus function, under natural light. For the experiments, a 13.3--inch display with 1280 x 800 resolution and 50\% screen brightness was used.

To examine the readability of the QR-code, we used three scanning distances, corresponding to the proper ergonomic position that a user views a smartphone (``Intimate distance''), a computer (``Personal distance'') and an ATM (``Social distance'') screen. Table~\ref{table-notations} provides specific information on these distances. For these three scenarios, we define three different units of data (208, 816, 1920 bits), based on the traditional Version 2 QR-code capacity that consists of 25 modules (26 characters), the Version 6 QR-code capacity used for long URL's (like Google Maps address URL's) (102 characters), and the Version 10 QR-code capacity that contain vCard contact information (240 characters). 

To create the QR-code image, it was highly important to correctly specify its physical size, so we relied on a mathematical formula~\cite{qr_size} to calculate the recommended minimum size based on the chosen information data and the distance for this experiment. ​

\textit{Minimum QR--code Size = (Scanning Distance / Distance Factor) * Data Density Factor}

\begin{itemize}
\item Distance Factor - Start off with a factor of 10 then reduce it by 1 for each of poor lighting in the scan environment. In our case we use 10 as the distance factor.
\item Data Density Factor: Counts the number of columns of dots in the QR--code image and then divides that by 25 to normalize it back to the equivalent of a Version 2 QR--code.

\end{itemize}

Table~\ref{table-results} summarizes the QR--code readability tests' results for the three different scanning distances, the three different information data units and the taken images under two different scanning angles ( 0$^{\circ}$ and 45$^{\circ}$). During the experiments, the display and the Google Glass were stable in the aforementioned distances. Thus we decided to perform the experiments under two different angles, replicating in this way two more realistic scenarios; i) the user is looking straight the display, or ii) she is looking under a small horizontal angle. We calculated the (``\textit{Accuracy}'') of the system, based on the successful QR--code decodes, after having performed for each case study 11 scans of the same QR--code.

Taking into consideration the evaluation results, we can argue that the \pname was able to recognize the QR--code and decode it with high accuracy most of the times. Still, in some cases, and in contrary to the results from the ``minimum QR--code size'' formula, \pname failed to detect or decode the QR--code. Failing to do so with the first try, it means that the user needs to activate again \pname in the one-time authentication mode or to wait T seconds (say, 5 (sec)) of an automated try in the continuous authentication mode. This failure happens due to either the small physical QR--code size or the low data density stored on it. Definitely, the accuracy of the system could be higher if the camera lens were better or the decode library was in position to decode the QR--code with a higher error correction level.

\begin{table}
\caption{Notations used in QR--code experiments}
\begin{center} 
      \scalebox{0.90}{ 
\begin{tabular}{  p{4cm}  p{4cm}  }
\hline

Intimate distance: &	20 / 7.87 (cm / inch) \\

Personal distance & 50 / 19.68 (cm / inch)  \\

Social distance & 120 / 47.2 (cm / inch) \\

Encoded bits  & 208 (28 characters) \\
  & 816 (102 characters)  \\

  &  1920 (240 characters) \\

\hline

\end{tabular}
}
\end{center}\label{table-notations}
\end{table}

\begin{table}
\caption{Stressing QR--codes}
\begin{center} 
      \scalebox{0.90}{ 

\begin{minipage}{2 in}
\begin{tabular}{ |l|l|l| }
\hline

\multicolumn{3}{ |c| }{Scanning angle 0$^{\circ}$} \\
\hline
Encoded bits & Distance & Accuracy \\ \hline
 \hline
\multirow{3}{*}{208} 
	& Intimate & 72.7\% \\
 	& Personal & 72.7\% \\
 	& Social & 100\% \\ \hline
\multirow{3}{*}{816} 
	& Intimate & 100\% \\
 	& Personal & 90.9\% \\
 	& Social & 100\% \\ \hline
\multirow{3}{*}{1920} 
	& Intimate & 54.5\% \\
 	& Personal & 100\% \\
  	& Social & 100\% \\
\hline
\end{tabular}
\end{minipage}
\begin{minipage}{2in}
 \begin{tabular}{ |l|l|l| }
\hline

\multicolumn{3}{ |c| }{Scanning angle 45$^{\circ}$} \\
\hline
Encoded bits & Distance & Accuracy \\ \hline
 \hline
\multirow{3}{*}{208} 
	& Intimate & 18.2\% \\
 	& Personal & 63.6\% \\
 	& Social & 100\% \\ \hline
\multirow{3}{*}{816} 
	& Intimate & 72.7\% \\
 	& Personal & 81.8\% \\
 	& Public & 100\% \\ \hline
\multirow{3}{*}{1920} 
	& Intimate & 27.3\% \\
 	& Personal & 81.8\% \\
  	& Social & 100\% \\
\hline
\end{tabular}  
\end{minipage}
}
\end{center}\label{table-results}
\end{table}

\begin{figure*}[tb]
    \centering
    \subfigure[75x75 pixels.]{
        \includegraphics[width=0.62\columnwidth]{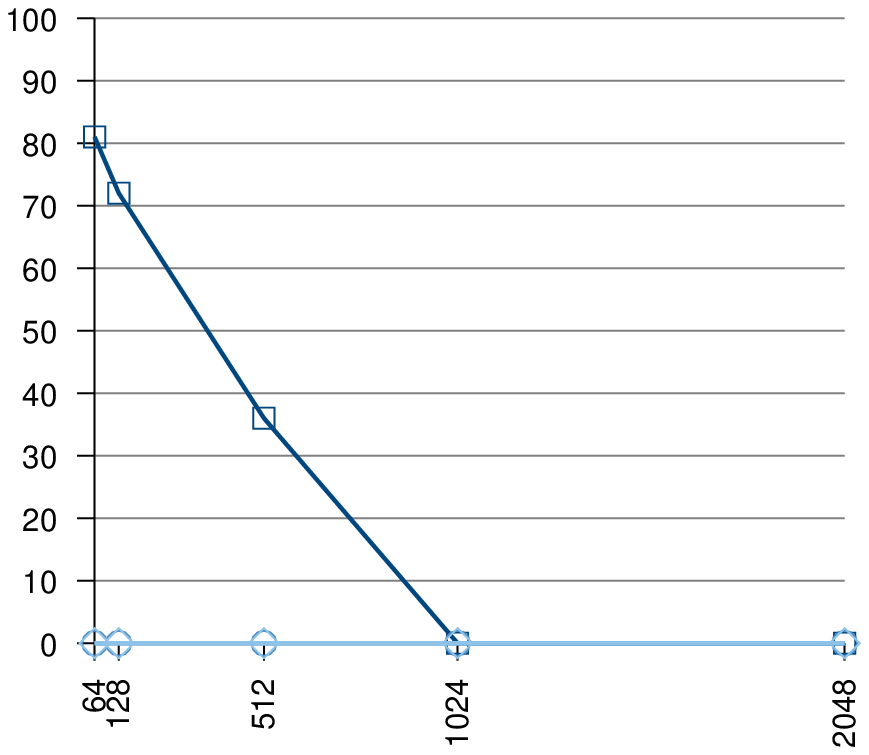}
        \label{fig:subfig1}
    }
    \subfigure[124x124 pixels.]{
        \includegraphics[width=0.62\columnwidth]{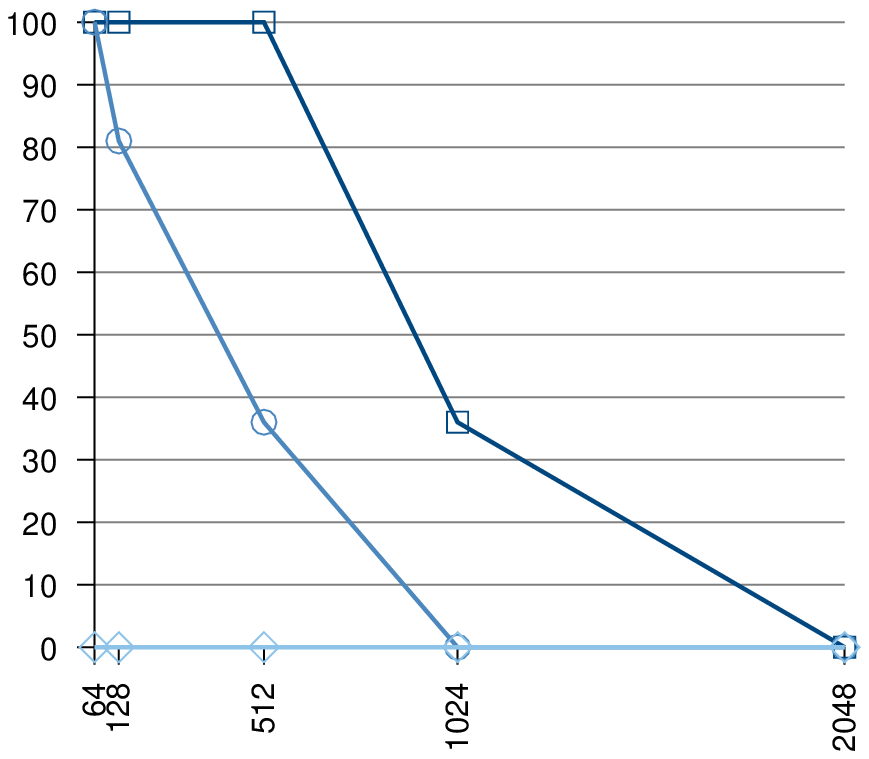}       
        \label{fig:subfig2}
    }
    \subfigure[170x170 pixels.]{
        \includegraphics[width=0.62\columnwidth]{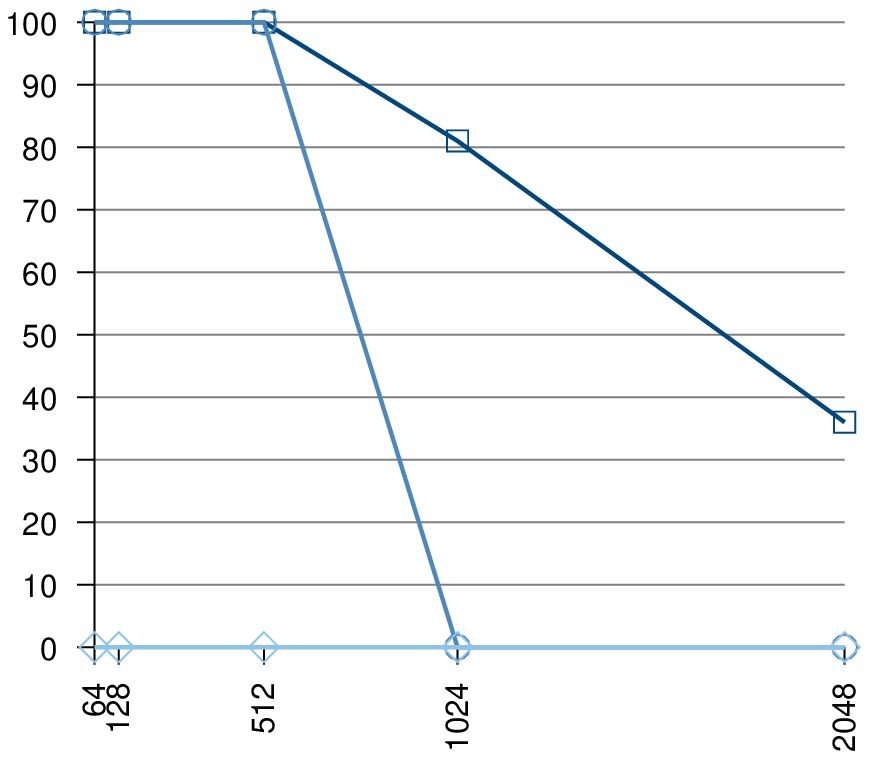}    
        \label{fig:subfig3}
    }
    \subfigure[188x188 pixels.]{
        \includegraphics[width=0.62\columnwidth]{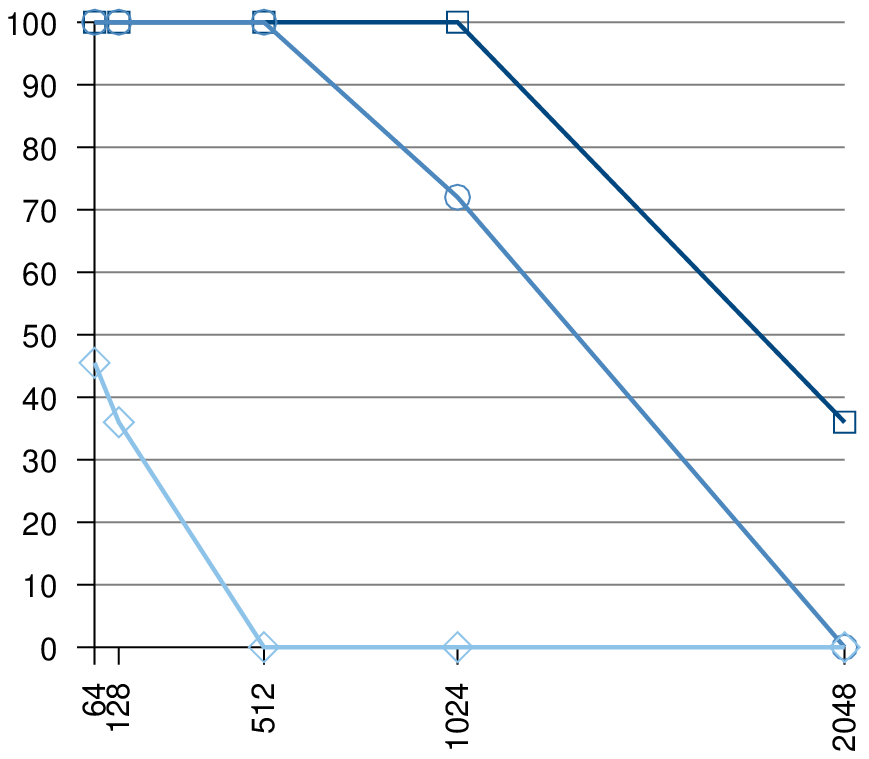}    
        \label{fig:subfig4}
    }
    \subfigure[309x309 pixels.]{
        \includegraphics[width=0.62\columnwidth]{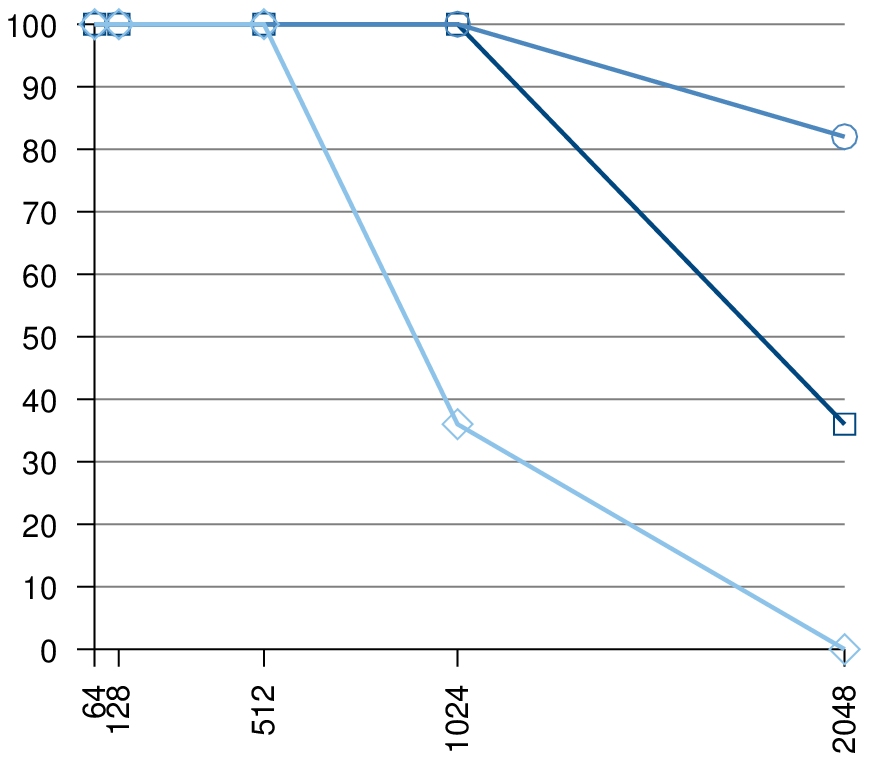}    
        \label{fig:subfig5}
    }
    \subfigure[430x430 pixels.]{
        \includegraphics[width=0.62\columnwidth]{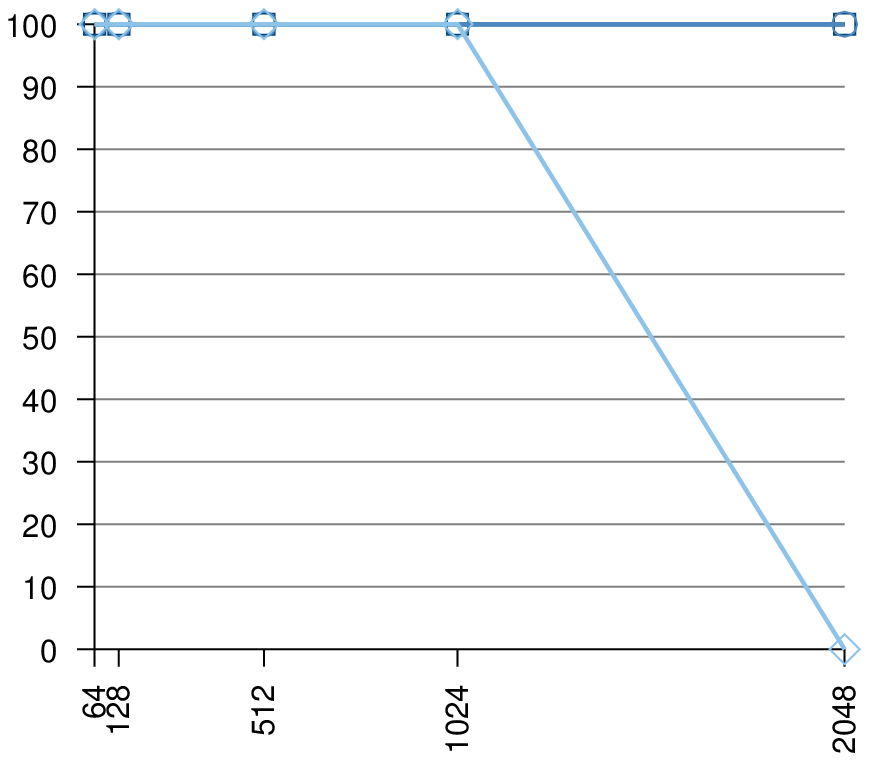}    
        \label{fig:subfig6}
    }
    \subfigure[453x453 pixels.]{
        \includegraphics[width=0.62\columnwidth]{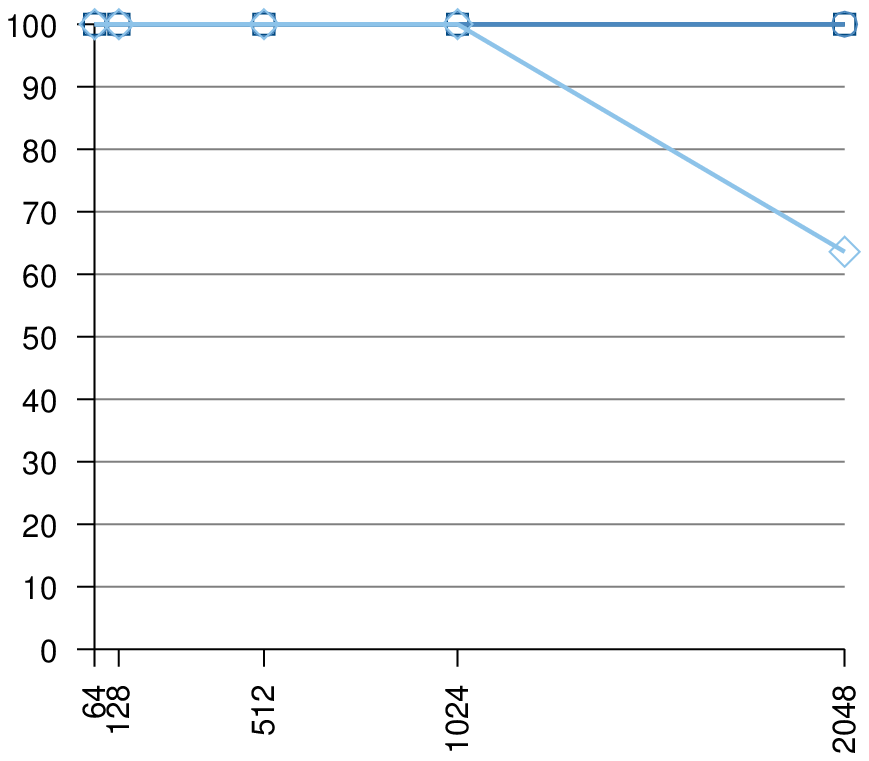}    
        \label{fig:subfig7}
    }
    \subfigure[744x744 pixels.]{
        \includegraphics[width=0.62\columnwidth]{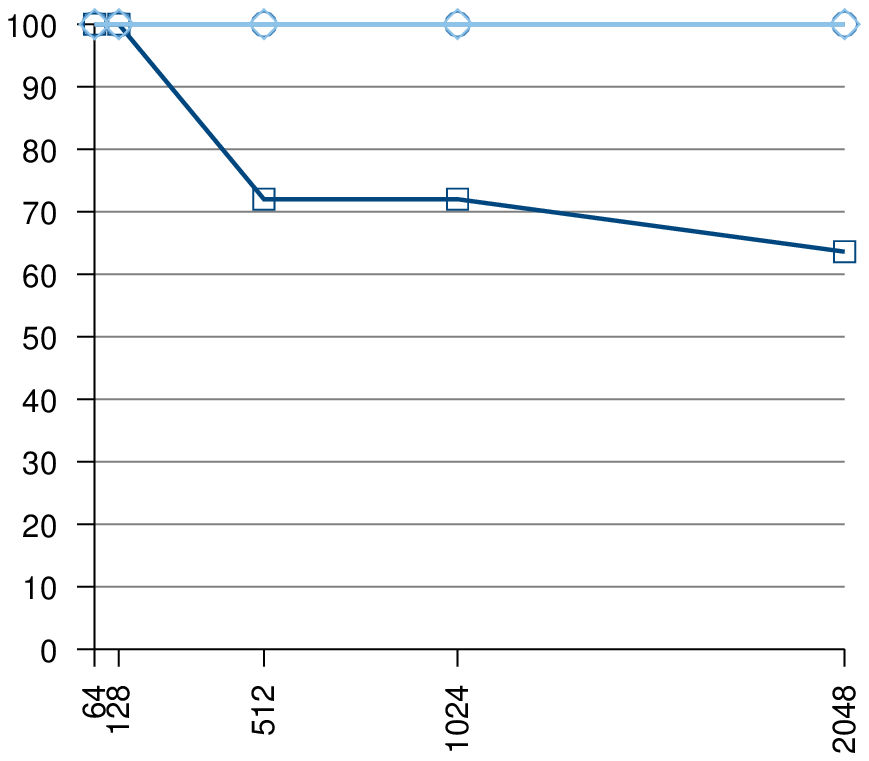}    
        \label{fig:subfig8}
    }
    \subfigure[1031x1031 pixels.]{
        \includegraphics[width=0.62\columnwidth]{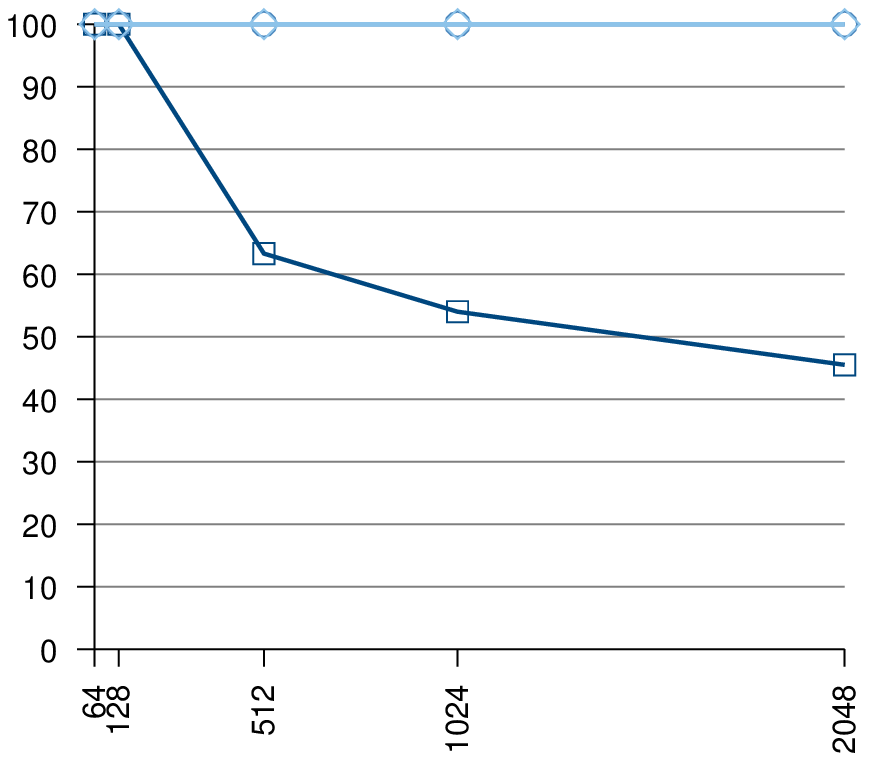}    
        \label{fig:subfig9}
    }
    \caption{Scanning different QR--code sizes from different 
    distances and no angle. Each sub--figure draws the percentage of correctly
    scanned QR 
    codes (\emph{y--axis}), when taking 11 pictures with Google Glass, as the 
    number of bits encoded in a QR--code increases (\emph{x--axis}). 
    The different colors correspond to three distances 
    between the glass and screen: 
    \crule[qr20]{6pt}{6pt} 20~cm,
    \crule[qr50]{6pt}{6pt} 50~cm, and 
    \crule[qr120]{6pt}{6pt} 120~cm. From sub--figure \ref{fig:subfig1}
    to \ref{fig:subfig9}, we increase the physical size of the QR--code.
    }
    \label{fig:QRlimits}
\end{figure*}

Before closing the QR--code readability chapter, we scan different QR--code sizes from different distances and no angle (refer to Fig.~\ref{fig:QRlimits}). Each sub--figure draws the percentage of correctly scanned QR--codes (y–axis) when taking 11 pictures with Google Glass. As the number of bits encoded in a QR--code increases (x–axissub–figure 7(a) to 7(i), we increase the physical size of the QR--code. The different colors correspond to three distances between the glass and screen: ￼ 20 cm, ￼50 cm, and ￼ 120 cm. From sub--figure 7(a) to 7(i), we increase the physical size of the QR--code. 
 
In Fig.\ref{fig:subfig1}, it is evident that we are able to correctly scan a QR--code only from a short distance of the screen. As the size of the code increases, we are in charge of scanning from longer distances. In Fig.~\ref{fig:subfig5}, the results for 20~cm overlap with the ones for 50~cm, thus they are not visible. For QR--codes larger than 453 pixels, we are always in power to correctly process them (Figures~\ref{fig:subfig8}--\ref{fig:subfig9}), even from a distance of 120~cm. ​

\subsection{User Study}

Designing a new human centric authentication scheme, like \pname, that can also be applicable by users with disabilities or user with the need to operate on multiple terminals, like the nurses, requires evaluation and feedback from real users through a pilot study. ​

A pilot study is a trial, which is conducted before the main study, and not only allows the researcher to define any problems and provide the required adjustments to the system, but also ensures whether or not the study is appropriate in terms of validity. 

For the pilot study, we recruited 20 volunteers to use a custom e--mail client, running on a laptop terminal, authenticating themselves, via the 5 authentication scenarios, to an email service. 

The duration of the total experiment was an hour per participant, and the precondition was that users should be computer, smartphone and e--mail users. All users had previous interaction with a smartphone, but none of them with a Google Glass. It is important to mention that none of these 20 volunteers was a multiple terminal operator or had a disability. For the experiments, proper IRB approval was granted, while users' privacy and sensitive information were never exposed. Table~\ref{table_demographics} contains a summary of our participants' demographic characteristics.​

\begin{table}
\caption{Participant demographics}
\begin{center} 
\begin{tabular}{ p{4cm} r  }
\hline
Characteristics & Total \\
\hline\hline
Gender & \\

Male		&	14 \\
Female	&	6	\\
\hline
Field & \\
\hline\hline
Computer Science (CS) &	12\\
Non-CS		&	8\\
\hline
Age & \\
\hline\hline
18-26 	&		11\\
26-32 	&		6\\
32-62	&		3\\
\hline
\end{tabular}
\end{center}
\label{table_demographics}
\end{table}

\subsubsection{Methodology}

During the experimentation phase, the participants had to use a custom e--mail client, running on a laptop terminal, authenticating themselves, via the 5 authentication scenarios, to an email Gauth--enabled service.
We should note that all users shared the same credentials during their single--factor authentication phase, while they had each time to activate the devices (smartphone and Glass) in order to perform the second factor authentication. ​

Due to the fact that none of the volunteers had previous experience with the Google Glass or our custom e-mail client, a pre--study tuition during the first 15 minutes of the experiment was mandatory. During that specific time, the instructor introduced the main functions of every hardware device that will be used during the experiment, including the Google Glass, and passed through them several times for every step of the experiment.
 
All five authentication scenarios share a common ruse in which participants were not told which authentication scheme was to be primary investigated. Each scenarios needed to be performed 3 times by all users, allowing us to micro--benchmark the authentication tasks and collect real data used during the evaluation phase.

At the end of the session, users had to complete an anonymized questionnaire survey related to their experience with the authentication scenarios. Additionally to the experimental phase and the questionnaire survey, a 5 minute conversation allowed the investigator to extract valuable information regarding the participants' thoughts.

 \subsubsection{Scenarios}

During the experimentation phase, the participants had to utilize the five authentication methods in order to access an e-mail account through a laptop terminal. First, we tested single--factor authentication. Users employed \pname and conventional username/password credentials to authenticate with the mock service. Second, we asked them to use a username/password as well as a second factor of authentication. We evaluated three scenarios: using \pname (2FA \pname), google--authenticator on a smartphone (2FA Smartphone), and google--authenticator on Google Glass (\mbox{2FA Glass}). In more details these scenarios include:

\begin{description}

\item[Username/Password]\hfill \\ the traditional single--factor authentication scheme based on username and password credentials.
\item[2FA -- Smartphone] \hfill \\a second factor authentication scheme, where a OTP was displayed by a mobile. 
\item[2FA--Glass]\hfill \\a second factor authentication scheme, where a OTP was displayed by the Google Glass.
\item[Gauth]\hfill \\ a password--less, hands--free, one--time authentication scheme 
\item[2FA--Gauth]\hfill \\the traditional singlest factor authentication scheme based on username and password credentials, working in synergy with \pname as the second authentication factor.

\end{description}

We should note that all users shared the same credentials during their single--factor authentication phase, while each time they had to activate the devices (smartphone and Glass) to perform the second--factor authentication.    
   
  \subsubsection{Results}

\begin{figure}
    \centering
    \includegraphics[width=\linewidth]{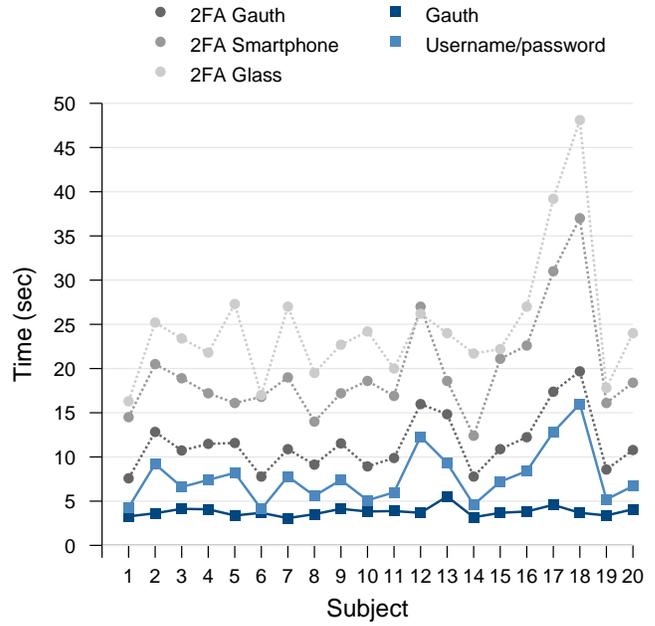}
    \caption{Time required to complete entry--point authentication for all 20 users in our user study. The figure draws the average over three repetitions of the task. First, we tested single--factor authentication. Users employed \pname and conventional username/password credentials to authenticate with a mock service (bottom two lines in the figure).  Second, we asked them to use a username/password as well as a second factor of authentication. We evaluated three scenarios: using \pname (2FA \pname), google--authenticator on a smartphone (2FA Smartphone), and google--authenticator on Google Glass (\mbox{2FA Glass}). }
    \label{fig:user_study}
\end{figure}

The Fig.~\ref{fig:user_study} illustrates the average over three repetitions of the task for all the five scenarios. The bottom two lines in the figure correspond to the single--factor authentication, while the rest three to the second factor authentication schemes. It is obvious that \pname, due to the hand-free nature, is not only the faster method in all cases, but grants novice users with tremendous authentication issues, like 17tn and 18tn user, to successfully authenticate in less than 5 seconds.​

Most of the users agreed that traditional authentication schemes need to be improved. As they comment during the oral discussion, password-less schemes like the \pname or the very recently iPhone's fingerprint solution permits them to remember less passwords. The 95\% argued that the combination of username/password with \pname, is the most secure authentication scheme across all the scenarios, but in their everyday life, users would choose \pname for its password-less nature. On the contrary, only 20\% feels that \pname is secure. This can be explained due to the limited or non cyber-security knowledge of our scheme. Most users, 65\%, complain that they had problem reading the OTP directly from the Glass, but claim that it is more secure than displaying the OTP on a smartphone. Last but not least, 75\% of the users allege that they are interested in using \pname on their everyday basis only if the cost of the wearable is affordable and more elegant, while a 65\% assert that they will use it only under specific circumstances, such as at their office, a bank terminal or to access a room.​





\subsection{Security}

\begin{figure}
    \centering
    \includegraphics[width=\linewidth]{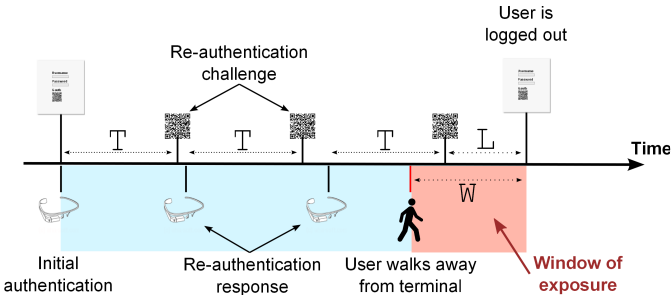}
    \caption{Window of exposure to malicious users when using \pname for 
    continuous authentication. $T$ is the re-authentication period used 
    (\ie how frequently it is requested), $L$ is the timeout (since the last 
    expected re--authentication) after which the user is logged out, and $W$ 
    the window of exposure.}
    \label{fig:window}
\end{figure}

When performing continuous authentication with period $T$, if a user steps 
away from the terminal without logging out, there exists a window of 
vulnerability $W$ that the terminal is left exposed, until the service 
detects that the user is no longer there and locks the terminal and logs 
the user out. This is illustrated in Fig.~\ref{fig:window}, where a user 
steps away, missing the next re-authentication request. Since the terminal 
probably does not immediately lock the terminal, an amount of time $L$ 
lapses till the user is logged out and the terminal is safe. The figure 
makes it easy to deduce that $W \leq T + L$. Since even 
with our current prototype, we are able to use a period of $T=5sec$. Also, 
since the authentication time on the local network never exceeds 0.5~sec, 
continuous authentication could be deployed with an $L\approx1sec$. 
Consequently, $W$ will be considerable small in most cases. However, future 
work is required to establish the usability effects of using such a small
$L$.
\section{Related Work}
\label{sec:related}

Over the last few years, various authentication methods and schemes have been introduced aiming to improving the security of computers and smartphone systems. Still passwords remain the most popular authentication scheme, although their weaknesses and limitations have been wildly explored. Users are often advised or required to choose passwords that comply with certain policies in order to strength their passwords, but in most cases they end-up with passwords easier to guess than they had previously assumed\cite{telepathwords:sec14}. 

As weak-password prevention and password managers schemes try to harden the traditional password, many promising schemes have been introduced over the last few year aiming to provide secure, cheap, and efficient authentication methods based on QR-codes, NFC tags or biometric features. So, the rest of this section will be categorize into three groups, presenting the most resent authentication schemes in the literature.

\subsection{Password Evaluation Studies}

Despite the fact that passwords suffer from various attacks, solutions like Telepathwords \cite{telepathwords:sec14} try to measures the password strength by simulating password-cracking algorithms and providing to the users strong passwords ~\cite{guessagain:oakland12}. Moreover, with the increasing hardware availability and the equipped devices with touchscreens and touch-surfaces, modern password allow user to draw a shape connecting some software points on a touch sensor as their password. These authentication schemes are referred as graphical passwords and aim to combine the usability of drawing a password with high guessability and entropy. Still graphical passwords suffer from smudge and shoulder surfing attacks~\cite{Aviv:2010,Tzong2014}. 
Measuring password strength by simulating password-cracking
algorithms~\cite{guessagain:oakland12}.

\subsection{Password Managers}

The increasing number of password users have to memorize, made passwords managers very popular over the last few years. Although the concept of managing password is not something new~\cite{Halderman:2005}, password managers have evolved from traditional software running in the browser, to cloud-based and mobile device based managers~\cite{Lackey2014,Silver2014}. While the the usability of password managers has been proven~\cite{Karole:2010}, various attacks have expose their vulnerabilities~\cite{Silver2014,Zhiwei2014}. In spite of the fact that, the concept of password managers is very promising, it cannot work on in many cases like on public terminals or embedded devices were password managers are not offered. 

%

\subsection{NFC-based authentication schemes}

NFC gains in popularity after smartphones got powered with NFC readers, allowing them to establish two-way contactless radio communication between endpoints by touching them together. NFC as authentication token has been widely utilized by various studies throughout the literature. Recent studies have proposed mobile credit card payments \cite{Choi2014}, health systems \cite{Dehling2014}, password-less authentication and identity management architectures used for securing web services \cite{poplett2014user, Mora2014, Memmi2014} .

\subsection{Authentication based on biometric}

While biometric is gaining in popularity the last few years, interesting solutions have been proposed in the resend literature trying to improve both two-factor and continuous authentication systems. The work in \cite{Frank2013} investigates the possible to continuous authenticate users while they perform basic navigation steps on a touchscreen device. Same year, a new way to perform continuous authentication using mouse dynamics as the behavioural biometric modality was demostrated in \cite{Mondal2013}. Very recently in \cite{Menshawy2014}, a general approach for continuous authentication based on keystroke dynamics was introduced. The authors prove, that it is feasible to authenticate users based on keystroke dynamics for continuous authentication systems. Also in \cite{Kambourakis2014}, a touchstroke dynamic system was introduced as a second verification factor when authenticating the user of a smartphone.

A pulse-response biometric~\cite{pulse-response:ndss14} sends an electric 
pulse through the user's body and measures is resistance. The expectation is 
that each user has a unique resistance which can be used to uniquely identify 
them to perform continuous authentication. However, a user's electric 
resistance can change based on various factors, measurements can be affected 
by weather conditions, while significant investments may be required to 
deploy such sensors on terminals. Also, is it not applicable to generic 
online services, similarly to the previous system. 

\subsection{Recurring authentication}

Recurring authentication using wearable devices like bracelets that keep track
of a user's hand movements and correlates the movement with the keys pressed to
continuously authenticate users \cite{zebra:oakland14}. They show 85\% 
accuracy and 11~sec adversary detection time, or 90\% and 50~sec. The error 
is too high and this approach only works on terminals, while it is harder to 
easily deploy as an additional authentication measure on generic online 
services offered through apps or the browser. Also, Google Glass is potentially
a multipurpose tool, like the smartphone, and is more likely to be adopted by 
a larger number of users, while a device can be actually shared by multiple
users.

Transparent re-authentication for mobile devices using behavioral
biometrics and in particular how a user interacts with the smartphone through
its touchscreen~\cite{reauthentication:ndss13}. False positives are disruptive.
Transparently authenticating users based on how they place
calls~\cite{howyouanswerme:asiaccs11}.

\subsection{Authentication for People with Disabilities}

Despite the majority of authentication schemes provided in the literature, people with disabilities have to come across once again with the limited authentication solutions provided in the literature. 
Authentication schemes based on multi-touch surfaces with with one or more fingers~\cite{passchords:assets12}, signal-based methods used to delineate the ECG features and determine the dominant fiducials from each heartbeat~\cite{ECGauthentication:icreate08}, eye-gaze gestures~\cite{eyegaze:ozchi07} are some interesting approaches thay may help and improve peoples every day authentication.


\section{Discussion}
\label{sec:discussion}

As already mentioned, continuous authentication with wearable devices is a relatively new discipline. Combining modern devices with security systems will not only improve the security of the system, but at the same time increase the usability and productivity of some end-user categories. 

\subsection{Device Theft}
\label{ssec:security}

Overall, with the increasing risk of the theft or loss of mobile devices over the last few years, authentication solutions like \pname may face extra issues. The person that holds the device, is also the one that can authenticate.
We can argue that the security of the glass device is similar to locking your password
vault. A PIN, voice recognition, or other techniques could be used to unlock
the device. The device could perform a challenge-response protocol with the user 
\cite{voicepass:way14, dontlisten:assets09}. For example, the device could 
display one character on the screen and the user would respond (by voice) by 
adding his secret number to the displayed one. Furthermore, blending \pname with physiological biometric schemes that rely on unique voice patterns, such as voice-based authentication systems, will allow only the legit user to control the device~\cite{Adibi2014137}.

\subsection{Privacy Benefits}

\pname preserves end-user privacy by shielding the user from keystroke profiling. For example, the service will not be able to infer what is the disability of the person authenticating, using methods like timing errors on a keyboard or other sensor data. Furthermore, pseudonymity can be provided on a per session basis for user that share the same device under their duty, like nurses or police officer. That is, each time the device creates a new unique OTP without requiring from the user extra credentials.

%
%
%

%
%
%
%
%
%
%
%

\section{Conclusion}
\label{sec:conclusions}

We presented \pname, a system that incorporates glass wearable devices in 
authentication. Our design enables the hands-free authentication of users 
with terminals, which we argue can have a great impact on how persons with 
disabilities currently access public terminals and shared devices in the 
workplace. We believe our system has the potential to empower people to 
overcome some of the physical barriers they face with regards to accessing 
online services. Equally important, \pname can help organizations safe 
guard their terminals from unauthorized access, incurred by the fact that 
users forget or avoid to log out from terminals they use. \pname allows 
services and terminals to determine  in seconds whether the correct user is 
operating them. We evaluated our approach by performing a short user study 
involving students. While the study does not include persons with 
disabilities, nor we evaluate continuous authentication in a real 
workplace, it is an important first step that has provided us with 
encouraging results. We plan to use these to pursue trials of \pname with 
users with disabilities and larger groups.

\bibliographystyle{IEEEtran}
\bibliography{main,sauth,portokalidis,ddamop}


\appendix
\subsection{Exit Survey}

The following questionnaire was given to the participants of the user study 
after the experiments. The number on the right of each response indicates 
how many participants chose it.

\begin{enumerate}
    \item Personal information
    \begin{enumerate}
        \item Are you a Male (M) or a Female (F)?\\
        \textcircled{ } Male     \surveyres{14}\\ 
        \textcircled{ } Female       \surveyres{16}
        
        \item How old are you?\\
        \textcircled{ } 18 - 26	 \surveyres{11}\\ 
        \textcircled{ } 26 - 32	 \surveyres{6}\\ 
        \textcircled{ } 32 +     \surveyres{3}\\
        
        \item Field of study\\
        \textcircled{ } Computer Science   \surveyres{12}\\ 
        \textcircled{ } Other              \surveyres{8}      
    \end{enumerate}
    
    \item General questions related to authentication methods
        \begin{enumerate}
        \item Which type of authentication method are you using in 
        your everyday life?\\
        \textcircled{ } Username/Password               \surveyres{20}\\
        \textcircled{ } Biometric                       \surveyres{3}\\
        \textcircled{ } Username/One-time password      \surveyres{4}\\
        \textcircled{ } Public/shared key cryptography  \surveyres{6}\\
        
        \item How secure are the following authentication methods?
        (4 = very secure, 3 = secure, 2 = basic security, 1 = not secure)\\
        \textcircled{ } Username/Password	             \surveyres{2}\\
        \textcircled{ } Biometric	                     \surveyres{2}\\
        \textcircled{ } Username/One-time password       \surveyres{3}\\
        \textcircled{ } Public/shared key cryptography   \surveyres{4}\\

        \item Do you feel that authentication methods should be improved
        in the future?\\
        \textcircled{ } Yes	                     \surveyres{17}\\
        \textcircled{ } No	                     \surveyres{3}\\
    \end{enumerate}
    
    \item Authentication through wearable devices
        \begin{enumerate}
        \item Is this the first time you are using a wearable device like 
        Google Glass?\\
        \textcircled{ } Yes                      \surveyres{20}\\
        \textcircled{ } No, but not mine         \surveyres{0}\\
        \textcircled{ } No, I already have one   \surveyres{0}\\
        
        \item Are you going to use wearable devices in your everyday life in 
        the near future?\\
        \textcircled{ } Yes	                     \surveyres{0}\\
        \textcircled{ } No	                     \surveyres{20}\\
        \textcircled{ } Only some                \surveyres{0}\\

        \item Do wearable devices improve the authentication process?\\
        \textcircled{ } Yes, both security and usability         \surveyres{13}\\
        \textcircled{ } No                                       \surveyres{2}\\
        \textcircled{ } Yes, but only usability, not security    \surveyres{5}\\
    \end{enumerate}
    
    \item Improving authentication via \pname
        \begin{enumerate}
        \item Which of the following scenarios do you feel is more secure?
        (Select only 1)\\
        \textcircled{ } Username/password                            \surveyres{0}\\
        \textcircled{ } \pname                                       \surveyres{0}\\
        \textcircled{ } Two-factor authentication with Smartphone    \surveyres{1}\\
        \textcircled{ } Two-factor authentication with Glass         \surveyres{0}\\
        \textcircled{ } Two-factor authentication with \pname        \surveyres{19}\\
    
        \item Do you think that \pname can improve the authentication process?
        \gp{Switched Google Glass for \pname here, because i have the feeling
        this was what was conveyed to the subjects}\\
        \textcircled{ } Yes, both security and usability         \surveyres{4}\\
        \textcircled{ } No                                       \surveyres{7}\\
        \textcircled{ } Yes, but only usability, not security    \surveyres{9}\\
        
        \item Which authentication method do you prefer?
        (Select only 1)\\
        \textcircled{ } Username/password                            \surveyres{0}\\
        \textcircled{ } \pname                                       \surveyres{0}\\
        \textcircled{ } Two-factor authentication with Smartphone    \surveyres{1}\\
        \textcircled{ } Two-factor authentication with Glass         \surveyres{0}\\
        \textcircled{ } Two-factor authentication with \pname        \surveyres{19}\\
        
        \item How secure are the following authentication methods?
        (4 = very secure, 3 = secure, 2 = basic security, 1 = not secure)\\
        \textcircled{ } Username/password                            \surveyres{2}\\
        \textcircled{ } \pname                                       \surveyres{2}\\
        \textcircled{ } Two-factor authentication with Smartphone    \surveyres{3}\\
        \textcircled{ } Two-factor authentication with Glass         \surveyres{3}\\
        \textcircled{ } Two-factor authentication with \pname        \surveyres{4}\\

        \item Would you use \pname in your everyday life?\\
        \textcircled{ } Yes	                     \surveyres{17}\\
        \textcircled{ } No	                     \surveyres{3}\\
    \end{enumerate}
\end{enumerate}

\end{document}